\documentclass[11pt]{article}
\usepackage{slashed,amsfonts}
\def\rmi{{\,\rm i\,}}
\newcommand{\Gl}{\mathop{\rm {}G}\ell }
\begin{document}
\thispagestyle{empty}
\begin{flushright}
DFTT 28/2004\\
hep-th/0412039\\[5cm]
\end{flushright}
\begin{center}
\textsc{\LARGE New aspects of $\mathcal N=2$ theories}\\[2cm]
Jos Gheerardyn\\[.5cm]
\emph{Dipartimento di Fisica Teorica, Universit\`a di Torino\\ and I.N.F.N. - Sezione di Torino,\\
Via P. Giuria 1, I-10125 Torino, Italy}\\[.3cm]
email: \texttt{gheerard@to.infn.it}\\[6cm]
\textbf{abstract}\\[.5cm]
\end{center}
We review some recent observations in theories with eight supercharges. We point out that such theories can be generalized by starting from the equations of motion rather than from an action. We show that vector multiplets constructed in such a way can have more general Fayet-Iliopoulos terms and that the scalar fields of hypermultiplets can be coordinate functions on more general target spaces. Although our discussion holds in five dimensions, the results can easily be extended to other dimensions.
\newpage
\section{Introduction}
During the past decade, $\mathcal N=2$ theories have proven to be a rich subject.  
The first reason why these theories are interesting is because of their simple but non-trivial structure. Although the superalgebra of symmetries imposes many constraints, there is still enough freedom to continuously deform the coupling functions. Otherwise stated, we can deform the target-space geometries while preserving the supersymmetry. On the other hand, classical field theories with more supercharges are completely fixed once the number and type of multiplets is known while theories with less supersymmetry are evidently less constrained and have therefore a far more intricate structure. 

Because of this close relationship between symmetry and structure, theories with eight supercharges have been used in a variety of fields. One of the most important applications is the research on dualities, initiated by the Seiberg-Witten papers \cite{Seiberg:1994rs,Seiberg:1994aj}. Another very interesting observation is that these theories can be constructed by considering string theory on a Calabi-Yau manifold. Through this connection, the geometry of these manifolds is mapped to quantities in four dimensional $\mathcal N=2$ supergravities \cite{Candelas:1990pi} leading to so-called special geometry. A similar connection for $\mathcal N=2$ compactifications on manifolds with flux remains to be enravelled. $\mathcal N=2$ theories were also considered in the context of AdS/CFT correspondence \cite{Aharony:1999ti}  and of brane world scenarios \cite{Bergshoeff:2000zn}.

Because of the many applications it is important to know the most general theory that is compatible with an $\mathcal N=2$ superalgebra. We will study so-called on-shell multiplets, which are multiplets that only realize the algebra correctly when they are put on the mass-shell. Hence, a fruitful way to obtain the most general theory is to directly study the algebra of infinitesimal transformations of the fields \cite{Bergshoeff:2002qk,Gheerardyn:2003rf} since demanding the closure of the algebra then yields all constraints, including the equations of motion. It has been shown that the field equations obtained in this way are not necessarily derivable from an action,\footnote{that is supersymmetric, Poincar\'e invariant and that has the same field content} hence they constitute more general theories that are compatible with the given supersymmetry.

Theories that do not admit an action are regularly encountered \cite{Gheerardyn:2004bz} since many different obstructions to a Lagrangian exist. The prime example is the existence of a self-dual tensor field, like in type IIB supergravity \cite{Schwarz:1983qr}. Another obstruction is that the scalars of the theory can be described by a non-linear sigma model with a target-manifold that does not allow for an invariant scalar product \cite{Bergshoeff:2002qk,Gheerardyn:2003rf,Stelle:2003rr}. The hypermultiplets that we will discuss in the present letter are examples of the latter. Yet another notable fact is that Scherk-Schwarz compactification of a theory with an action might yield one without. There exists a massive type IIA supergravity (not equivalent to Roman's massive type IIA) that does not admit a Lagrangian \cite{Howe:1997qt,Lavrinenko:1997qa,Gheerardyn:2002wp} and that can be constructed in this way. Note also that demanding the vanishing of beta functions in string theory and solving torsional constraints in superspace both yield equations of motion directly. Hence, these techniques might or might not lead to equations of motion that are integrable into an action. Let us finally point out that the theories we will describe could have e.g. a non-supersymmetric action or a Lagrangian containing auxiliary fields. This would render these theories even more intriguing.

The content of this letter is as follows. In Section \ref{vec} we will discuss on-shell vector multiplets in rigid supersymmetry and we will show the existence of new Fayet-Iliopoulos terms. In Section \ref{hyp} we will hold a similar discussion for hypermultiplets which will lead to the discovery of new target manifolds. The final Section \ref{disc} contains our conclusions and some open problems.
\section{On-shell vector multiplets}\label{vec}
The physical fields of an on-shell vector multiplet in five dimensions are a real scalar $\sigma$, a real vector $A_\mu$ that gauges a local ${\rm U}(1)$ symmetry and two fermions $\psi^i$, with $i=1,2$, that transform as a doublet under the ${\rm SU}(2)$ R-symmetry. If we have $n$ multiplets, we will label the fields with an extra index $I=1,\dots,n$ and in that case, the $n$ vectors $A_\mu^ I$ will gauge an $n$ dimensional Lie-group where the corresponding Lie-algebra has structure constants $f_{IJ}{}^K$ and coupling constant $g$ and both the scalars and the fermions will transform in the adjoint representation of this group.

Using dimensional analysis, one can argue that the most general supersymmetry transformations of these fields are
\begin{eqnarray}\label{susytrans}
\delta_Q \sigma^I&=&\frac12 \rmi \bar \epsilon \psi^I\,,\quad \delta_Q A_\mu^I=\frac12 \bar \epsilon \gamma_\mu \psi^I\,,\nonumber\\
\delta_Q \psi^{iI}&=&-\frac12 \rmi \slashed{\mathfrak D} \sigma^I \epsilon^i-\frac14 \slashed F^I \epsilon^i+g f^{(ij)I}\epsilon_j-\frac12 \rmi \gamma^ I_{JK} \bar \psi^{iJ}\psi^{jK}\epsilon_j\,.
\end{eqnarray}
Note that $\epsilon^i$ are the parameters of the supersymmetry, that we do not write ${\rm SU}(2)$ indices on singlet fermion-bilinears and that 
\begin{equation}
F_{\mu\nu}^I=2\partial_{[\mu}A_{\nu]}^I+g f_{JK}{}^IA_\mu^JA_\nu^K\,,\quad \mathfrak D_\mu \sigma^I=\partial_\mu \sigma^I+gf_{JK}{}^I A_\mu^J \sigma^K\,.
\end{equation}
In equation (\ref{susytrans}) the only unknowns are the objects $f^{(ij)I}$ and $\gamma_{JK}{}^ I$ which are at this point general functionals of the scalar fields and where $\gamma^ I_{JK}$ is symmetric in the lower indices.

From \cite{Gunaydin:1983bi}, it is known that the commutator of two supersymmetries yields a translation $P$ and a field dependent gauge transformation $G$
\begin{equation}\label{algv}
\lbrack \delta_{Q}(\epsilon_1),\delta_{Q}(\epsilon_2)\rbrack=\delta_P(\frac12 \bar \epsilon_2 \gamma^\mu \epsilon_1)+\delta_G(-\frac12 \rmi \bar \epsilon_2 \epsilon_1)\,.
\end{equation}
Computing this commutator for the bosonic fields, one finds the desired result (\ref{algv}) while a similar computation for the fermions yields
\begin{eqnarray}
\lbrack \delta_Q(\epsilon_1),\delta_Q(\epsilon_2)\rbrack\psi^{iI}&=&\delta_P(\frac12 \bar \epsilon_2 \gamma^\mu \epsilon_1)\psi^{iI}+\delta_G(-\frac12 \rmi \bar \epsilon_2 \epsilon_1)\psi^{iI}\nonumber\\&&-\frac{1}{16}(3 \bar \epsilon_2\epsilon_1\Gamma^{iI}+3\bar \epsilon_2 \gamma^ \mu \epsilon_1 \gamma_\mu\Gamma^{iI}+\bar \epsilon_2^{(i}\gamma^{\mu\nu}\epsilon_1^{j)} \gamma_{\mu\nu} \Gamma^I_j)\,,\nonumber\\
\Gamma^{iI}&=&\slashed{\mathfrak{D}}\psi^{iI}+\gamma^I_{JK}\slashed{\mathfrak{D}}\sigma^J\psi^{iK} +\frac12\rmi\gamma^I_{JK}\slashed{F}^J\psi^{iK}-\frac12 \partial_K\gamma_{JL}^I \bar{\psi}^{iJ}\psi^{jL}\psi^K_j\nonumber\\&&+2\rmi g\gamma^I_{JK}f^{ijI}\psi_j^K +\rmi gf_{JK}{}^I\sigma^J\psi^{iK}\,,\label{GI}
\end{eqnarray}
together with the first two conditions listed in Table \ref{tab:1}. The fact that a supersymmetry has to commute with a gauge transformation then yields the other conditions mentioned in that Table. 
\begin{table}[tb]
\caption{Conditions imposed by symmetries}
\label{tab:1}\renewcommand{\arraystretch}{1.5}
\begin{tabular}{l|ll} 
 & $\lbrack \delta_Q,\delta_Q\rbrack=\delta_P+\delta_G$ & $\lbrack\delta_Q,\delta_G\rbrack=0$\\ \hline
 $\gamma^I_{JK}$&$\gamma^I_{LM}\gamma^M_{JK}=-\frac12 \partial_L \gamma_{JK}^I$&$2
 f_{J(L}{}^M\gamma_{K)M}^I-f_{JM}{}^I\gamma^M_{KL}+f_{JM}{}^N\sigma^M\partial_N\gamma^I_{KL}=0$\\
$f^{(ij)I}$ & $\partial_Jf^{(ij)I}+2\gamma^I_{JK}f^{(ij)K}=0 $ & $f_{JL}{}^K\sigma^L\partial_Kf^{(ij)I}-f_{JK}{}^If^{(ij)K}=0$ \\
\end{tabular}
\end{table}

Since we want the closure of the algebra, we deduce from (\ref{GI}) that $\Gamma^{iI}\equiv 0$, hence we have found the equation of motion for the fermions. Moreover, this dynamical constraint forms a multiplet with the equations of motion for the bosonic fields, which read
\begin{eqnarray}
\Box \sigma^I+ \gamma_{JK}^I\mathfrak{D}_\mu \sigma^J\mathfrak{D}^\mu \sigma^K-\frac12 \gamma^I_{JK}F_{\mu \nu}^JF^{\mu \nu K}+2g^2\gamma^I_{JK}f^{ijJ}f_{ij}^K&=&0\,,\nonumber\\
\mathfrak{D}^\nu F_{\nu \mu}^I-\frac14\gamma^I_{JK}\varepsilon_{\mu \nu \rho \sigma \tau}F^{\nu\rho J} F^{\sigma \tau K} +2 \gamma^I_{JK}\mathfrak{D}^\nu \sigma^KF_{\nu \mu}^J-gf_{JK}{}^I\sigma^J\mathfrak{D}_\mu \sigma^K&=&0\,,\label{EOM}
\end{eqnarray}
up to fermionic terms. Note first of all that the $\mathcal N=2$ dynamics are completely encoded in the functionals $\gamma$ and $f$ which should satisfy the constraints of Table \ref{tab:1}. The latter objects $f$ should be compared to the Fayet-Iliopoulos terms in the more familiar off-shell vector multiplet. This can be seen from the potential term in the scalar equation of motion (\ref{EOM}) which is similar to the FI potential in the off-shell case. However, in the case under consideration, the FI terms are more general and even possible in the non-Abelian case. The object $\gamma$ is to be compared to the completely symmetric gauge-invariant three tensor $C_{IJK}$ in the off-shell formulation. But again, the present case is more general.

Using the Batalin-Vilkovisky formalism \cite{Batalin:1984jr}, we have moreover determined precisely what the extra conditions are to be able to write down a supersymmetric, Poincar\'e invariant action that has the same field content. It has turned out \cite{Gheerardyn:2003rf} that the object $\gamma$ then corresponds to the Levi-Civita connection computed from the very special metric $g_{IJ}=C_{IJK}\sigma^K$ while only constant FI terms are possible in the Abelian factors of the gauge group.

Let us illustrate the above with the following example. Consider two Abelian vector multiplets. This implies that only the conditions in the first column of Table \ref{tab:1} are non-trivial. We can take
\begin{equation}
\gamma^1_{11}=\frac1{2\sigma^1}\,,\quad \gamma^1_{22}=\frac1{\sigma^1\sigma^2}\,,\quad \gamma^2_{22}=\frac1{2\sigma^2}\,,\quad f^1=\frac2{\sigma^1\sigma^2}\,,\quad f^2=\frac1{\sigma^2}\,,
\end{equation}
and the other components of $\gamma$ equal to zero. These objects determine the dynamics of a theory that cannot admit an action since it is easy to show that there cannot be any metric that yields this $\gamma$ as its Levi-Civita connection.
\section{On-shell hypermultiplets}\label{hyp}
We can use the same method for the hypermultiplet, which consists of four real scalars $q$ and two fermions $\zeta$. We will again consider $n$ multiplets together. The most general rigid supersymmetry transformations read \cite{Bergshoeff:2002qk}
\begin{equation}\label{susyq}
\delta_Q q^X=-\rmi \bar \epsilon^i\zeta^A f_{iA}^X\,,\quad \delta_Q \zeta^ A=\frac12\rmi \slashed \partial q^ X f_X^{iA} \epsilon_i-\zeta^B \omega_{XB}{}^A \delta_Q q^X\,.
\end{equation}
The index $A$ runs from $1$ to $2n$, $X=1,\dots,4n$, and the objects $f_X^{iA}$, $f_{iA}^X$ and $\omega_{XA}{}^B$ are at this point unknown functionals of the scalars. Similar to the previous case we can now find conditions on these functionals by considering the superalgebra. Computing this first on the scalars, we find that the commutator of two supersymmetries only yields a translation when 
\begin{equation}\label{covcte}
f_Y^{iA}f_{iA}^X=\delta_Y^X\,,\quad f_X^{iA}f_{jB}^Y=\delta_j^i\delta_B^A\,,\quad \mathfrak D_X f^Y_{iA}\equiv\partial_Xf_{iA}^Y-\omega_{XA}{}^Bf_{iB}^Y+\Gamma^Y_{XZ}f^Z_{iA}=0\,,
\end{equation}
where $\Gamma$ is a functional of the scalar fields and is symmetric in its lower indices. Since the dynamics of hypermultiplets is governed by a non-linear sigma model, the scalars should be interpreted as coordinate functions on some target space. This means that our formulation should be covariant under redefinitions $q'{}^X=q'{}^X(q)$ and $\zeta'{}^A=\zeta^B U_B{}^A$ where $U$ is a $\Gl(n,\mathbb H)$ matrix that may depend on the scalars.\footnote{The fact that $U$ takes values in $\Gl(n,\mathbb H)$ is a consequence of the reality conditions that have to be imposed \cite{Bergshoeff:2002qk}.} Under the first set of reparametrisations $f_X^{iA}$ and $\omega$ transform as one-forms, $f^X_{iA}$ as a vector and $\Gamma$ as a connection. Under the second set of redefintions, $\omega$ transforms as a connection. These observations imply that $f_X^{iA}$ is a Vielbein, $f^X_{iA}$ its inverse, $\Gamma$ a torsionless connection with holonomy contained in $\Gl(n,\mathbb H)$ and $\omega$ a spin-connection.

Hence, we can construct a triple of tensors
\begin{equation}\label{defJ}
\vec J {}_X{}^Y=-\rmi f_X^{iA}\vec \sigma_i{}^j f_{jA}^Y\,,
\end{equation}
where $\vec \sigma$ are the three Pauli matrices $\sigma^\alpha$, $\alpha=1,2,3$ written as a vector. These tensors generate the quaternionic algebra
\begin{equation}
J^\alpha{}_X{}^Z J^\beta{}_Z{}^Y=-\delta^{\alpha \beta} \delta_X^Y  +\varepsilon^{\alpha\beta\gamma}J^\gamma{}_X{}^Y\,,
\end{equation}
and are covariantly constant with respect to the torsionless connection $\Gamma$ by virtue of (\ref{covcte}). Such a set of tensors is called a hypercomplex structure, and the corresponding manifold a hypercomplex manifold. The existence of this structure implies that the structure group of the frame bundle can be reduced to $\Gl(n,\mathbb H)$ as we already discovered above. Finally, note that if one computes the commutator of two supersymmetries on the fermions, this not only yields a translation but also non-closure functionals similar to (\ref{GI}) which become equations of motion upon closure of the algebra. In conclusion, starting from the infinitesimal transformations (\ref{susyq}), we have found that rigid hypermultiplets can parametrise a hypercomplex manifold. We have also shown how to find the corresponding field equations that govern the dynamics.

If one requires that these equations of motion are derivable from an action, one is forced to introduce a metric $g$ on the target manifold. This can easily be seen from the fact that the kinetic term for the scalars should look like
\begin{equation}
\mathcal L_{kin}=-\frac12 g_{XY}\partial_\mu q^X\partial^\mu q^Y\,.
\end{equation}
Note that the necessity of a metric can again rigorously be proven by the use of the Batalin-Vilkovisky formalism. Invariance of the action then demands that the metric is covariantly constant with respect to the connection $\Gamma$ and that the metric is tri-hermitian, i.e.
\begin{equation}
\vec J_X{}^Zg_{YZ}=-\vec J_Y{}^Z g_{XZ}\,.
\end{equation}
In conclusion, an action is only possible if the target manifold is hyperk\"ahler \cite{Alvarez-Gaume:1981hm}.

We can also demand that a conformal symmetry is realized \cite{Bergshoeff:2002qk}. We then have to introduce a dilatation transformation, which looks like
\begin{equation}
\delta_D q^X=\Lambda k^X\,,
\end{equation}
where $\Lambda$ is the infinitesimal parameter and $k$ is a vector on the hypercomplex space. Algebraic considerations then lead to the fact that
\begin{equation}
\mathfrak D_X k^Y=\frac32 \delta_X^Y\,.
\end{equation}
Using this vector, we can construct three other vectorfields that generate an ${\rm SU}(2)$ action on the manifold in the following way:
\begin{equation}
\vec k^X=\frac13 \vec J_Y{}^X k^Y\,.
\end{equation}
The important conclusion here is that the curvature, which is the generator of the holonomy, leaves these four vectorfields invariant, i.e.
\begin{equation}
\mathfrak D_X \mathfrak D_Y k^Z=0 \Rightarrow R_{XVY}{}^Z k^V=R_{XVY}{}^Z \vec k^V=0\,.
\end{equation}
Hence, the holonomy fixes one quaternion and should therefore be contained in ${\rm SU}(2)\cdot \Gl(n-1,\mathbb H)$. Otherwise stated, a conformal hypercomplex manifold has holonomy contained in this group.

We can repeat the above in the context of local $\mathcal N=2$ supersymmetry as well. Starting from the most general supersymmetry transformations \cite{Ceresole:2000jd,Bergshoeff:2004kh}
\begin{equation}
\delta_Q q^X=-\rmi \bar \epsilon^i\zeta^A f_{iA}^X\,,\quad \delta_Q \zeta^ A=\frac12\rmi \slashed \partial q^ X f_X^{iA} \epsilon_i-\zeta^B \omega_{XB}{}^A-\frac12 \bar\psi_\mu^i\zeta^A \gamma^\mu \epsilon_i\,,
\end{equation}
where $\psi$ denotes the gravitino, we can again compute the commutator of two supersymmetries. It now turns out that on the right hand side, not only a translation but also a field dependent supersymmetry transformation appears, which deforms the covariant constancy of the Vielbeine to
\begin{equation}
\mathfrak D_X f^Y_{iA}\equiv\partial_Xf_{iA}^Y-\omega_{XA}{}^Bf_{iB}^Y-\omega_{Xi}{}^jf_{jB}^Y+\Gamma^Y_{XZ}f^Z_{iA}=0\,,
\end{equation}
where $\omega_{Xi}{}^j$ is a non-trivial ${\rm SU}(2)$ connection and $\Gamma$ is again torsionless. This implies that the holonomy of $\Gamma$ now is contained in ${\rm SU}(2)\cdot \Gl(n,\mathbb H)$, for a $4n$ dimensional target manifold. Introducting now again the triplet of tensors $\vec J$, as in (\ref{defJ}), we see that these tensors now are covariantly constant with respect to the affine connection $\Gamma$ and the ${\rm SU}(2)$ connection. Such a set of tensors is called a quaternionic structure and the corresponding manifold a quaternionic manifold. 

As above, the equations of motion could be computed from the superalgebra realized on the fermions. If we demand the existence of an action, we have again to introduce a tri-hermitian metric $g$ and the target manifold becomes quaternionic-K\"ahler \cite{Bagger:1983tt}.

Due to the fact that the holonomy groups of a $4(n+1)$ dimensional conformal hypercomplex and a $4n$ dimensional quaternionic manifold coincide, there exists a close relationship between these manifolds \cite{Swann,PedersenPS1998,Bergshoeff:2004nf} since the space of leaves, generated by the four vector fields $k$ and $\vec k$ on the conformal hypercomplex manifold is a quaternionic manifold. This observation has physical consequences since it can be used in the superconformal tensor calculus approach (see e.g. \cite{VanProeyen:1983wk} for a review) to the construction of new $\mathcal N=2$ supergravities that do not admit an action.
\section{Discussion}\label{disc}
We have pointed out that it is possible to find more general $\mathcal N=2$ theories by restricting to equations of motion rather than to a least action principle. In the case of vector multiplets this leads to more general Fayet-Iliopoulos terms and hence, more general potentials. In the case of hypermultiplets, we have found new target spaces. 

It would be very interesting to find the string theoretic origin of such theories. Therefore, the next step in the discussion of these theories should be the search for an appropriate compactification Ansatz for which the above theories would be the low-energy truncation. If such a compactification scheme exists, we would naturally be lead to reconsider the $\mathcal N=2$ applications to see how these theories would yield new results.
\section*{Acknowledgements}
First of all, I would like to thank the organisers of the RTN-EXT workshop 2004, to give me the opportunity to present this work and to choose Kolymbari as the conference venue. I would also like to thank my collaborators Eric Bergshoeff, Sorin Cucu, Tim de Wit, Stefan Vandoren and Toine Van Proeyen. This work has been supported in part by the F.W.O.-Vlaanderen to which the author is affiliated as postdoctoral researcher, by the Italian M.I.U.R. under the contract P.R.I.N. 2003023852, ``Physics of fundamental interactions: gauge theories, gravity and strings'' and by the European Community's Human
Potential Programme under contract MRTN-CT-2004-005104 `Constituents,
fundamental forces and symmetries of the universe'.

\end{document}